\documentclass[twocolumn,aps,floats,floatfix,prl,superscriptaddress,tightenlines]{revtex4}
\usepackage{graphicx}
\usepackage{bm}
\usepackage{epsfig}
\usepackage{pstricks}
\usepackage{amsmath}
\usepackage{hyperref}

\begin{document}

\title{Next-to-leading QCD effect to the quark compositeness search at the LHC}
\author{Jun Gao}
\author{Chong Sheng Li}\email{csli@pku.edu.cn}
\author{Jian Wang}
\author{Hua Xing Zhu}
\affiliation{Department of Physics and State Key Laboratory of Nuclear Physics
and Technology, Peking University, Beijing 100871, China}
\author{C.-P.Yuan}\email{yuan@pa.msu.edu}
\affiliation{Department of Physics and Astronomy, Michigan State University,
East Lansing, 48824, USA}

\pacs{12.38.Bx,~12.60.Rc,~13.87.-a}

\begin{abstract}
We present the exact next-to-leading order (NLO) QCD corrections to the dijet production
induced by the quark contact interactions at the CERN Large Hadron Collider (LHC). We
show that as compared to the exact calculation, the scaled NLO QCD prediction adopted by
the ATLAS Collaboration has overestimated the new physics effect on some direct
observables by more than 30\% and renders a higher limit on the quark compositeness
scale. The destructive contribution from the exact NLO correction will also lower the
compositeness scale limit set by the CMS Collaboration.
\end{abstract}

\maketitle

The quark composite models have been studied extensively in the
literature~\cite{Eichten:1983hw,Lane:1996gr}. It is assumed that quarks are
composed of more fundamental particles with new strong interactions at a
composite scale $\Lambda$, much greater than the quark masses. At energy well
below $\Lambda$, quark contact interactions are induced by the underlying
strong dynamics, and yield observable signals at hadron colliders. For example,
the dijet production at the CERN Large Hadron Collider (LHC) could be largely
modified. In the Standard Model (SM), the theory of Quantum Chromodynamics
(QCD) predicts the jets in dijet events are preferably produced in large
rapidity region, via small angle scatterings in t-channel processes. On the
contrary, the dijet angular distribution induced by the quark contact
interactions is expected to be more isotropic. The D0 and CDF Collaborations at
the Fermilab Tevatron have set limits on the scale $\Lambda$ based on their
dijet data~\cite{Abe:1996mj}. Recently, both the ATLAS and CMS collaborations
have carried out similar
analyses~\cite{Collaboration:2010eza,Khachatryan:2010te} using the LHC dijet
data with $\sqrt{s}=7\ {\rm TeV}$, in proton-proton collisions,
and an integrated luminosity of about $3\
{\rm pb}^{-1}$. The observed limits for $\Lambda$ at the $95\%$ confidence
level (CL) are $3.4\ {\rm TeV}$ and $4.0\ {\rm TeV}$, respectively, which have
already exceeded the previous limits. With more integrated luminosity
collected, these limits will be further improved.

The limits on the composite scale $\Lambda$ were obtained by comparing the experimental
dijet data with various theory predictions, including the SM next-to-leading order (NLO)
QCD corrections and contribution induced by the quark contact interactions~which were
handled differently in different experiments. While the CMS Collaboration included only
the leading order (LO) contribution from the quark contact
interactions~\cite{Khachatryan:2010te}, the D0, CDF and ATLAS Collaborations included the
{}``scaled NLO QCD correction'' which assumes the NLO correction (in terms of K-factors)
to the dijet production from the contact interactions to be exactly the same as that from
the SM QCD interactions~\cite{Abe:1996mj,Collaboration:2010eza}. In this Letter, we
present the exact NLO QCD correction to the dijet production induced by the quark contact
interactions, and discuss its impact to the existing experimental limits set by both the
ATLAS and CMS Collaborations.

To compare with the experimental analyses, we consider only the quark contact
interactions that are the products of left-handed electroweak isoscalar quark
currents which are assumed to be flavor-symmetric to avoid large
flavor-changing neutral-current interactions~\cite{Lane:1996gr}. The effective
Lagrangian can be written as
\begin{equation}
\mathcal{L}_{NP}=\frac{1}{2\Lambda^{2}}(c_{1}O_{1}+c_{2}O_{2}),
\end{equation}
where $c_{1}$, $c_{2}$ are the Wilson coefficients, and $O_{1}$, $O_{2}$ are
the color-singlet and color-octet operators given by
\begin{eqnarray}
O_{1} & = & \delta_{ij}\delta_{kl}\left(\sum_{c=1}^{3}\bar{q}_{Lci}\gamma_{\mu}
q_{Lcj}\sum_{d=1}^{3}\bar{q}_{Ldk}\gamma^{\mu}q_{Ldl}\right),\nonumber \\
O_{2} & = & {\rm T}_{ij}^{a}{\rm T}_{kl}^{a}\left(\sum_{c=1}^{3}\bar{q}_{Lci}
\gamma_{\mu}q_{Lcj}\sum_{d=1}^{3}\bar{q}_{Ldk}\gamma^{\mu}q_{Ldl}\right),
\end{eqnarray}
in which $c$, $d$ are the generation indices and $i$, $j$, $k$, $l$, $a$ are the color
indices, and ${\rm T}^{a}$ are the Gell-Mann matrices. The Wilson coefficients at the
scale $\Lambda$ are conventionally normalized to be $c_{1}(\Lambda)=4\pi\cos{\theta},\
c_{2}(\Lambda)=4\pi\sin{\theta}$, with $0\leq\theta<2\pi$. Notice that the above
operators can also arise from the exchange of new heavy resonances in various new physics
models, such as $Z'$ models~\cite{Langacker:2008yv} and extra dimensions
models~\cite{Randall:1999ee}. Thus, our analyses are rather model independent and
$\Lambda$ can be identified as the effective new physics (NP) scale. When using the above
operators to calculate an observable at a scale much lower than $\Lambda$, we have to
consider the QCD running effects of the Wilson coefficients~\cite{Buchalla:1995vs}, which
can be easily derived by solving the renormalization group equation using the one-loop
anomalous dimension matrix of the operators, as following: \begin{equation}
\left(\begin{array}{c}
c_{1}(\mu_{R})\\
c_{2}(\mu_{R})\end{array}\right)=\left(\begin{array}{cc}
\frac{N+1}{2N} & \frac{N-1}{2N}\\
1 & -1\end{array}\right)\left(\begin{array}{c}
b_{1}r^{\frac{-3(N-1)}{N\beta_{0}}}\\
b_{2}r^{\frac{3(N+1)}{N\beta_{0}}}\end{array}\right),
\end{equation}
with $r=\alpha_{s}(\mu_{R})/\alpha_{s}(\Lambda)$, and
\begin{equation}
b_{1}=c_{1}(\Lambda)+\frac{C_{F}}{N+1}c_{2}(\Lambda),\
b_{2}=c_{1}(\Lambda)-\frac{C_{F}}{N-1}c_{2}(\Lambda),
\end{equation}
where $N=3$ and $C_{F}=4/3$ for QCD, $\beta_{0}=(11N-2n_{f})/3$, and $\mu_{R}$
is the renormalization scale, $n_{f}=5$ is the number of active quark flavors.
In the studies of the ATLAS and CMS Collaborations, they focus on the
color-singlet operator with destructive interference, which corresponds to
$\theta=0$ in our analyses. Below, we will first discuss this case, and then
extend to more general cases with arbitrary $\theta$ values.

At the LO, there are several subprocesses which contribute to the dijet
production at hadron colliders induced by the NP operators we considered,
including
\begin{equation}
qq'(q)\rightarrow qq'(q),\ q\bar{q}'\rightarrow
q\bar{q}',\ q\bar{q}\rightarrow q\bar{q}(q'\bar{q}'),
\end{equation}
where $q$, $q'$ could be all the quarks except the top quark. The NP contributions
included in our calculation consist of two parts, the NP squared terms and the
interference terms between the NP and the SM QCD interactions, which yield different
kinematic distributions. We carried out the NLO calculations in the Feynman-'t Hooft
gauge with dimensional regularization (DR) scheme (with naive $\gamma_5$
prescription)~\cite{Buchalla:1995vs} in $n=4-2\epsilon$ dimensions to regularize all the
divergences. Below, we only show the analytical results for the subprocess
$q(p_{1})q'(p_{2})\rightarrow q(p_{3})q'(p_{4})$, since the similar results for other
subprocesses can be obtained by crossing symmetry.

First, we define the following abbreviations for the color structures and the
matrix element,
\begin{eqnarray}
\mathcal{M}_{0} & = & \bar{u}_{L}(p_{3})\gamma_{\mu}u_{L}(p_{1})
\bar{u}_{L}(p_{4})\gamma^{\mu}u_{L}(p_{2}),\nonumber \\
\mathcal{C}_{1} & = & \delta_{i_{3}i_{1}}\delta_{i_{4}i_{2}},\
\mathcal{C}_{2}={\rm T}_{i_{3}i_{1}}^{a}{\rm T}_{i_{4}i_{2}}^{a},
\end{eqnarray}
where $i_{1-4}$ are the color indices of the external quarks. The LO scattering
amplitudes induced by the NP and the SM QCD interactions can be separately written as
\begin{eqnarray}
i\mathcal{M}_{NP}^{tree} & = & i\mathcal{M}_{0}(c_{1}\mathcal{C}_{1}
+c_{2}\mathcal{C}_{2})/\Lambda^{2},\nonumber \\
i\mathcal{M}_{SM}^{tree} & = &
i\mathcal{M}_{0}(4\pi\alpha_{s}\mathcal{C}_{2})/t,
\end{eqnarray}
where $s,\ t,\ u$ are the Mandelstam variables, and we only keep the
left-handed current product of the SM QCD amplitudes here since others have no
interference with the NP interactions. After adding the 1-loop amplitudes and
the corresponding counterterms, we have the ultraviolet finite virtual
amplitudes as follows. \allowdisplaybreaks{\begin{eqnarray}
i\mathcal{M}^v_{NP}&=&i\mathcal{M}_0C_{\epsilon}{\alpha_s\over 4\pi}
\Big\{-2C_F\Big[c_1\mathcal{A}
(t)+{c_2\over 2N}\mathcal{B}(u)\Big]\mathcal{C}_1\nonumber\\
&&\hspace{-1cm}+\Big[ c_1\big(-2\mathcal{B}(u)\big)+c_2\big(-2C_F
\mathcal{A}(u)+{1\over N}\big(\mathcal{B}(u)+\nonumber\\
&&\hspace{-1cm} \mathcal{B}(t)
\big)\big)\Big]\mathcal{C}_2\Big\}/\Lambda^2,\nonumber\\
i\mathcal{M}^v_{SM}&=&i\mathcal{M}_0C_{\epsilon}{\alpha_s\over 4\pi}
\Big\{4\pi\alpha_s\Big[-{C_F\over 2N}\Big({4\over \epsilon}\ln(-{s\over u})-
\nonumber\\
&&\hspace{-1cm}2{t\over s}\ln({t\over u})-{u^2\over s^2}\ln^2({t\over
u})+\ln^2({s^2 \over t u})+(1-{u^2\over
s^2})\pi^2\Big)\Big]\nonumber\\
&&\hspace{-1cm}\mathcal{C}_1+4\pi\alpha_s\Big[-2C_F\Big( {2\over
\epsilon^2}+{1\over\epsilon}\big(3+2\ln(-{s\over u})\big)\Big)\nonumber\\
&&\hspace{-1cm}+{2\over N\epsilon}\ln({s^2\over t u})
+\beta_0\ln({\mu_R^2\over s})-\big({2\over3}n_f-{10\over3}C_F-{8\over3N}\big)\nonumber\\
&&\hspace{-1cm}\ln(-{s\over t})+{3\over N}\ln^2(-{s\over t}) -\big({1\over
2N}-C_F\big)\Big({u^2\over s^2}\big(\ln^2({t \over
u})\nonumber\\
&&\hspace{-1cm}+\pi^2\big)-2{u\over s}\ln({t\over u})+\ln^2({t\over u})-
2\ln(-{s\over u})\big(1\nonumber+\\
&&\hspace{-1cm}\ln(-{s\over u})\big)\Big)
+\big(C_F+{3\over2N}\big)\pi^2-\big({10\over9}n_f-{26\over9}C_F\nonumber\\
&&\hspace{-1cm}-{85\over9N}\big) \Big]\mathcal{C}_2\Big\}/t, \label{e1}
\end{eqnarray}}
where $C_{\epsilon}=({4\pi\mu_R^2\over
s})^{\epsilon}{1\over\Gamma(1-\epsilon)}$, and
\begin{eqnarray}
\mathcal{B}(x)&=&{2\over \epsilon}\ln(-{s\over x})+3\ln(-{\mu_R^2\over x})
+\ln^2(-{s\over x})+\pi^2+9,\nonumber\\
\mathcal{A}(x)&=&{2\over \epsilon^2}+{3\over \epsilon}+({2\over \epsilon}+3)\ln(-{s\over
x})+\ln^2(-{s\over x})+8.
\end{eqnarray}
We have checked that the virtual correction for the SM QCD contributions given in
Eq.~(\ref{e1}) agrees with the ones shown in Ref.~\cite{Ellis:1985er}. The infrared
divergences in virtual corrections should cancel with those in real corrections. As for
the real corrections, we apply both the two cutoff method~\cite{Harris:2001sx} and the
dipole subtraction method~\cite{Catani:1996vz} in our calculations for a cross-check.
Furthermore, we do not include loop corrections which would induce other new high
dimension operators and involve more uncertainties, and will discuss them elsewhere.

To compare with the dijet measurements of the ATLAS and CMS, we adopt the
anti-$k_{T}$ jet algorithm~\cite{Cacciari:2008gp} with the $E_{T}$
recombination scheme~\cite{Bayatian:2006zz} at parton level. We use the same
parton distribution function sets~\cite{Pumplin:2002vw} and set both the
renormalization and factorization scales to the average jet transverse momentum
$p_{T}$, as in the experimental analyses. Since our calculations are based on
the effective field theory approach, we only consider the dijet invariant mas
$m_{jj}$ up to the NP scale $\Lambda$. The SM QCD contributions are also
calculated up to NLO, using both the modified EKS code~\cite{Ellis:1992en} and
the NLOjet++ program~\cite{Nagy:2003tz} for a cross-check.

\begin{figure}[h]
 \includegraphics[width=0.45\textwidth]{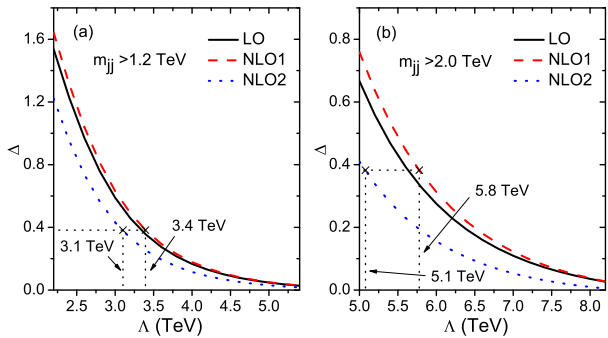}
\caption[]{$\Delta$ as functions of the scale $\Lambda$. The NLO1 (2) curves
represent the scaled (exact) NLO results.}
\label{f1}
\end{figure}

\begin{figure}[h]
\includegraphics[width=0.3\textwidth]{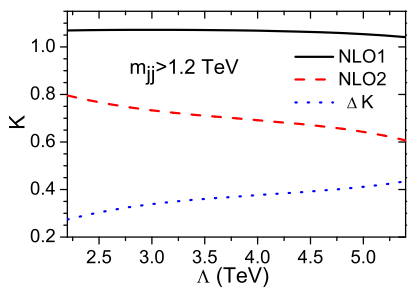}
\caption[]{The scaled NLO and the exact NLO K-factors, and their differences,
as functions of $\Lambda$.}
\label{f2}
\end{figure}

The ATLAS measurement --- Following the ATLAS analysis, we impose the following
kinematic cuts for selecting the dijet events. \begin{eqnarray}
|y_{jet}|<2.8,\ p_{T1(2)}>60(30){\rm GeV},\ m_{jj}>1.2{\rm TeV},\nonumber \\
|y_{b}|=|y_{1}+y_{2}|/2<0.75,\ |y^{*}|=|y_{1}-y_{2}|/2<1.7.\end{eqnarray} The jet radius
parameter $\Delta R$ is chosen to be 0.6. We note that at parton level, the constraints
on $y^{*}$ and $m_{jj}$ have already set a threshold of about 150 ${\rm GeV}$ for
$p_{T2}$. To quantify the NP effects on dijet angular distribution, we divide the region
of $|y^{*}|$, between 0 and 3.3/2, into 11 bins with equal bin width, and define $F$ as
the ratio of the number of events in the first 4 bins (with cross section $\sigma(4th)$)
to that in the total 11 bins (with cross section
$\sigma(tot)$)~\cite{Collaboration:2010eza}. In Fig.~\ref{f1}(a) we show the effect of
new physics contribution, defined through the discriminator
$\Delta=(F_{NP+SM}-F_{SM})/F_{SM}$, i.e.,
\begin{equation}
\Delta=\dfrac{F_{NP+SM}}{F_{SM}}-1=\frac{\sigma_{NP+SM}(4th)
/\sigma_{SM}(4th)}{\sigma_{NP+SM}(tot)/\sigma_{SM}(tot)}-1,
\end{equation}
as a function of the compositeness scale $\Lambda$ in three cases: (1)
including only the LO results as done by CMS; (2) using the scaled NLO QCD
results as done by ATLAS; (3) including the exact NLO QCD corrections presented
in this Letter. While the scaled NLO QCD calculation predicts a higher value of
$\Delta$ than the LO calculation, the exact NLO QCD calculation yields a much
smaller value. To further quantify the differences, we plot in Fig.~\ref{f2}
the ratios (K-factors) of the scaled NLO (labeled as NLO1) and the exact NLO
(labeled as NLO2) predictions for $\Delta$ to the LO results as functions of
$\Lambda$. The difference in the two NLO K-factors is also shown, which is
generally larger than 30\% for $\Lambda$ larger than about 3 TeV. Using the
scaled NLO prediction, ATLAS found the 95\% CL exclusion limit on the quark
compositeness scale is 3.4 ${\rm TeV}$, which corresponds to $\Delta$ value
about 0.38. If using the exact NLO QCD calculation, the exclusion limit will
reduce by about 10\% to 3.1 ${\rm TeV}$.

Fig.~\ref{f1}(b) shows similar results as Fig.~\ref{f1}(a), but requiring the
dijet invariant mass to be larger than 2 TeV instead of 1.2 TeV. This
measurement will become feasible when more integrated luminosity is collected
at the LHC. Again, the scaled NLO calculation would overestimate the exclusion
limit as compared to the exact NLO QCD calculation. For $\Delta=0.38$, the
difference is more than 10\%.

\begin{figure}[h]
\includegraphics[width=0.45\textwidth]{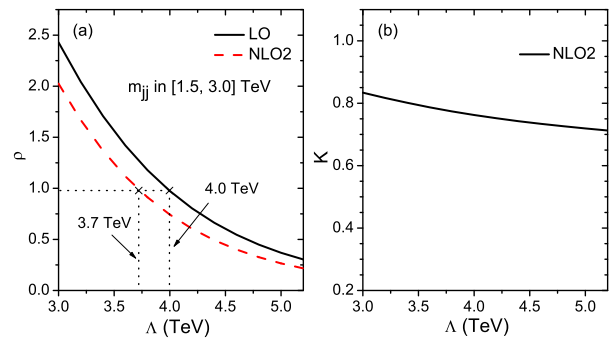}
\caption[]{(a) $\rho$ as functions of the compositeness scale $\Lambda$, as
predicted by the LO and the exact NLO calculations. (b) The exact
NLO K-factors of $\rho$, as functions of $\Lambda$. }
\label{f3}
\end{figure}

The CMS measurement --- Following the CMS analyses, we require
\begin{equation}
|\eta_{jet}|<1.3,\ p_{Tjet}>50{\rm GeV},\ \Delta R=0.7,
\end{equation}
and define the centrality ratio $R^{\eta}$ as the number of events in the inner
pseudo-rapidity region $|\eta_{1,2}|<0.7$ (with cross section $\sigma_{in}$) to
the one in the outer region $0.7<|\eta_{1,2}|<1.3$ (with cross section
$\sigma_{out}$) to quantify the NP effects on the dijet angular distribution,
and evaluate $R^{\eta}$ in each invariant mass bin from several hundred GeV to
3 TeV~\cite{Khachatryan:2010te}. In our calculation we use the same cuts and
strategy as the CMS except that we only consider the last five invariant mass
bins with equal width ranging from 1.5 TeV to 3 TeV, since the NP effects are
more significant in this region. To quantitatively compare various theory
predictions, we introduce the discriminator $\rho$ as the weighted average
deviation of $R^{\eta}$ from the SM predictions,
\begin{eqnarray}
\rho&=&\Big(\sum_{i=1}^{5}\omega_i{\big |}R^{\eta}_{NP+SM}(i)-R^{\eta}_{SM}(i)
{\big |}/R^{\eta}_{SM}(i)\Big)/
\sum_{i=1}^{5}\omega_i,\nonumber \\
\omega_i&=&\sqrt{{\mathcal L}/\big(\sigma_{in,NP+SM}(i)^{-1}+\sigma_{out,NP+SM}(i)^{-1}\big)},
\end{eqnarray}
where $\omega_{i}$ is the inverse of the relative statistical error of
$R_{NP+SM}^{\eta}(i)$, and ${\mathcal{L}}$ is the integrated luminosity. In
Fig.~\ref{f3}(a), we plot $\rho$ as functions of the compositeness scale
$\Lambda$, as predicted by the LO and the exact NLO calculations. The exact NLO
QCD corrections generally reduce the value of $\rho$ as compared to the LO
prediction. For completeness, the exact NLO K-factors of $\rho$ are shown in
Fig.~\ref{f3}(b). Consequently, the exact NLO QCD corrections modify the CMS
95\% CL exclusion limit of $\Lambda$ from 4.0 TeV to 3.7 TeV.

\begin{figure}[h]
\includegraphics[width=0.45\textwidth]{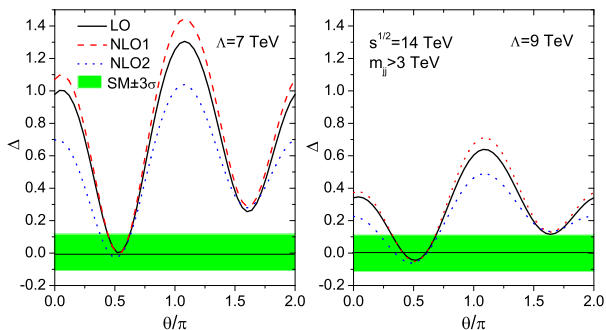}
\caption[]{$\Delta$ as functions of $\theta$ for different
values of $\Lambda$.}
\label{f4}
\end{figure}

General cases with arbitrary $\theta$ values --- It is interesting to also study the
general cases in which both the $c_{1}(\Lambda)$ and $c_{2}(\Lambda)$ coefficients in the
effective Lagrangian are nonzero. Here, we consider the discovery potential of the LHC
with $\sqrt{s}=14\ {\rm TeV}$, and raise the dijet invariant mass cut to 3 TeV. In
Fig.~\ref{f4} we plot the discriminator $\Delta$ as functions of $\theta$ for different
values of $\Lambda$ together with the 3$\sigma$ error bands from the SM predictions,
which are calculated by assuming an integrated luminosity of 1 ${\rm fb^{-1}}$. We find
that $\Delta$ varies largely from positive to negative values with the change of $\theta$
for a given $\Lambda$ value, and the difference between the scaled NLO and the exact NLO
predictions can be larger than 30\%. Another interesting observation is that there are
regions of theory parameter space in which the NP effects on the angular distribution
vanish, due to the cancellation of the contributions from the $O_{1}$ and $O_{2}$
operators. In that case, it will be difficult to study the quark compositeness scale
through the dijet angular distribution measurement at the LHC. Finally, the quark contact
interactions can also modify the top quark pair production at the Tevatron and the LHC.
It may be possible to distinguish the effects induced from the two operators by
studying the correlations between the dijet and the top quark pair
 production processes, which will be presented
elsewhere.

In conclusion, we have calculated the exact NLO QCD corrections to the dijet
production at the LHC, induced by the quark contact interactions which may
arise from the quark compositeness models or other new physics models. From our
results, the current exclusion limits of the quark compositeness scale set by
the ATLAS and CMS Collaborations shall be lowered from 3.4 TeV to 3.1 TeV, and
4.0 TeV to 3.7 TeV, respectively. Moreover, we discussed the general cases with
color-octet operator included in the quark contact interactions and found that
in some regions of the theory parameter space, the quark compositeness may
become undetectable at the LHC via the dijet angular distribution measurement.

\begin{acknowledgments}
This work was supported in part by the National Natural Science
Foundation of China, under Grants No.11021092 and No.10975004. C.P.Y
acknowledges the support of the U.S. National Science Foundation
under Grand No. PHY-0855561.
\end{acknowledgments}

\end{document}